\documentclass
[superscriptaddress,secnumarabic,amssymb,amsmath,nobibnotes,aps,prd,showkeys,showpacs,nofootinbib,12pt]{revtex4}%
\usepackage{graphicx}
\usepackage{epsf}
\usepackage{bm}
\usepackage{amsmath}
\usepackage{amsfonts}
\usepackage{amssymb}%
\providecommand{\U}[1]{\protect\rule{.1in}{.1in}}
\newcommand{\be}{\begin{equation}}
\newcommand{\ee}{\end{equation}}

\newcommand{\mincir}{\raise
-3.truept\hbox{\rlap{\hbox{$\sim$}}\raise4.truept\hbox{$<$}\ }}
\newcommand{\magcir}{\raise
-3.truept\hbox{\rlap{\hbox{$\sim$}}\raise4.truept\hbox{$>$}\ }}

\begin{document}
\title{Hyperbolic inflation in the Jordan frame}
\author{Andronikos Paliathanasis}
\email{anpaliat@phys.uoa.gr}
\affiliation{Institute of Systems Science, Durban University of Technology, Durban 4000,
South Africa }
\affiliation{Instituto de Ciencias F\'{\i}sicas y Matem\'{a}ticas, Universidad Austral de
Chile, Valdivia 5090000, Chile}

\begin{abstract}
We consider a multi-scalar field model in the Jordan frame which can be seen
as a two-scalar field model where the Brans-Dicke field interacts in the
kinetic part with the second scalar field. This theory under a conformal
transformation reduces to the hyperbolic inflation. We show that scaling
solutions and the de Sitter universe are provided by theory. In the study of
asymptotic dynamics, we determine an attractor where all the fluid sources
contribute in the cosmological fluid. This attractor is always a spiral and it
can be seen as the analogue of the hyperbolic inflation in the Jordan frame.

\end{abstract}
\maketitle

\section{Introduction}

\label{sec1}

Brand and Dicke proposed a scalar field gravitational theory which embodied
Mach's Principle \cite{BD}. In Brans-Dicke theory, the presence of the scalar
field is essential for the existence of the gravitational field, because the
scalar field is nonminimally coupled to gravity. That is, the gravitational
theory is defined in the Jordan frame \cite{Jord}. This gravitational theory
has been widely studied before for the description of various epochs of the
cosmological history, see for instance
\cite{ssen,Boi,Boi1,Boi2,Clif12,DEAmendola,omegaBDGR,BER,bsjd} and references
therein. Brans-Dicke theory belongs to the family of modified theories of
gravity and the scalar field can attribute dynamical degrees of freedom
provided by higher-order theories, such as in the case of the $f\left(
R\right)  $-theory \cite{Aref1}.

On the other hand, General Relativity is a theory defined in the so-called
Einstein frame. There exists a geometrical map which connects the Jordan frame
to the Einstein one and vice versa \cite{bb00a}. The two frames are related
under the Action of a conformal transformation \cite{bb00}. The two frames
have different metric tensors, which are conformally related, thus the
physical properties of a given theory in the two frames should be compared,
and the problem of the selection of the physical frame follows \cite{bb01}.
The latter has been a subject of debate in the last decades and there are
various studies on this analysis. The two frames are only mathematically
equivalent while the physical properties of a gravitational theory rarely
survives. For instance a singular solution in the one frame it can be read as
a solution without singularity in the other frame \cite{bb02,bb03,bb04}.

A gravitational model which has drawn the attention of cosmologists over the
last years is a multi-scalar field model in the Einstein frame which provides
the hyperbolic inflation \cite{ch5}. In this theory the inflaton field
\cite{guth}, which drives the dynamics of the rapid expansion of the universe,
consists by two scalar fields which interact in the kinetic components of the
Lagrangian function \cite{ch5}. Specifically, the dynamics of the two scalar
fields are defined on a two-dimensional hyperbolic space. This model is also
known as Chiral theory and it has been widely studied in the literature before
\cite{ch6,sco02,sch1,sch2,sch3,ch3,ch4,sco01,mm1,mm2}. This two-scalar field
model for the exponential scalar field potential it admits an attractor which
can provide the asymptotic behaviour of the hyperbolic inflation. In
particular it was found that this inflationary solution is described by a
stable spiral in terms of dynamical systems \cite{ch2}.

In this study we propose a new multi-scalar field model defined in the Jordan
frame and under the application of a conformal transformation it is equivalent
to the hyperbolic inflationary model, or the Chiral model. One of the scalar
fields is the Brans-Dicke field while the second scalar field is coupled to
the Brans-Dicke field only. The main idea of this study is to investigate if
there exists an analogue of the hyperbolic inflation in the Jordan frame and
to compare the physical properties of the two-scalar field model in the
Einstein and Jordan frames. In order to achieve our goal we apply methods from
the dynamical systems and we investigate the stationary points for the field
equations and we study their stability criteria. The dynamical analysis is an
important mathematical tool for the qualitative study of a gravitational
theory. It has been applied in various models in order to test the
cosmological viability of theories as also to reconstruct the cosmological
history \cite{dyn14,dn1,dn4,dn5,dn6,dn10,dn11}. The plan of the paper is as follows.

In Section \ref{sec2} we present the multi-field cosmological model of our
consideration and we derive the field equations in the case of a spatially
flat isotropic and homogeneous geometry. In Section \ref{sec4} we show the
existence of exact solutions of special interest; indeed, we prove that the
field equations can be solved explicitly and the scaling and de Sitter
solutions are provided by the proposed theory. The main analysis of this study
is presented in Section \ref{sec5}. In particular we perform a detailed
analysis of the dynamics of the field equations by using dimensional variables
in the Hubble normalization approach. We show that there exists an analogue of
the hyperbolic inflation in the Jordan frame. Finally, in Section \ref{conc}
we draw our conclusions and we extend our discussion in the case of
anisotropic spacetimes.

\section{Cosmological model}

\label{sec2}

Consider the multi-scalar field\ Action Integral%
\begin{equation}
S_{A}=\int dx^{4}\sqrt{-g}\left[  \frac{1}{2}\phi R+\frac{1}{2}\frac
{\omega_{BD}}{\phi}g^{\mu\nu}\phi_{;\mu}\phi_{;\nu}+\frac{1}{2}F^{2}\left(
\phi\right)  g^{\mu\nu}\psi_{;\mu}\psi_{;\nu}+V\left(  \phi,\psi\right)
\right]  , \label{bb.01}%
\end{equation}
where $\phi\left(  x^{k}\right)  $ and $\psi\left(  x^{k}\right)  $ are two
scalar fields which we assume that they inherit the symmetries of the
background space and in general they interact through the function $F\left(
\phi\right)  $ and the potential function $V\left(  \phi,\psi\right)  $.

Scalar field $\phi\left(  x^{k}\right)  $ is the Brans-Dicke scalar field
\cite{BD} coupled to gravity and $\omega_{BD}$ is the Brans-Dicke parameter
\cite{omegaBDGR}, with $\omega_{BD}\neq-\frac{3}{2}$. We remark that for
$\omega_{BD}=0$, the Brans-Dicke field describes the higher-order derivatives
of $f\left(  R\right)  $-gravity \cite{Aref1}, while for $\omega_{BD}%
=-\frac{3}{2}$ the Brans-Dicke field does not describe real degrees of freedom.

For the Action Integral (\ref{bb.01}) the gravitational field equations are
\begin{align}
\phi G_{\mu\nu}  &  =\frac{\omega_{BD}}{\phi}\left(  \phi_{;\mu}\phi_{;\nu
}-\frac{1}{2}g_{\mu\nu}g^{\kappa\lambda}\phi_{;\kappa}\phi_{;\lambda}\right)
-\left(  g_{\mu\nu}g^{\kappa\lambda}\phi_{;\kappa\lambda}-\phi_{;\mu}%
\phi_{;\nu}\right) \nonumber\\
&  +g_{\mu\nu}V\left(  \phi,\psi\right)  -F^{2}\left(  \phi\right)  \left(
\psi_{;\mu}\psi_{;\nu}-\frac{1}{2}g_{\mu\nu}g^{\kappa\lambda}\psi
_{;\kappa\lambda}\right)  ~, \label{bb.02}%
\end{align}
and the equations of motion for the two scalar fields
\begin{equation}
\left(  2\omega_{BD}+3\right)  \left(  g^{\mu\nu}\phi_{;\mu\nu}\right)  -\phi
V_{,\phi}+2V-\frac{1}{2}\left(  F^{2}\right)  _{,\phi}\dot{\psi}^{2}=0~,
\end{equation}%
\begin{equation}
g^{\mu\nu}\left(  \psi_{;\mu\nu}+\left(  \ln\left(  F^{2}\right)  \right)
_{,\phi}\phi_{;\mu}\psi_{;\nu}\right)  =0.
\end{equation}

The gravitational model (\ref{bb.01}) has not been defined as arbitrary. More
specifically, the multi-scalar field Action Integral (\ref{bb.01}) can be seen
as the equivalent of the hyperbolic inflation in the Jordan frame for a
specific functional form of $F\left(  \phi\right)  $. Indeed, when we perform
the conformal transformation $\bar{g}=\phi g_{ij}$ the Action Integral
(\ref{bb.01}) in the Einstein frame reads%

\begin{equation}
S_{A}=\int dx^{4}\sqrt{-g}\left[  \frac{1}{2}R-\frac{1}{2}g^{\mu\nu}\Phi
_{;\mu}\Phi_{;\nu}-\frac{1}{2}F^{2}\left(  \Phi\right)  e^{\bar{\omega}\Phi
}g^{\mu\nu}\psi_{;\mu}\psi_{;\nu}-\bar{V}\left(  \Phi,\psi\right)  \right]  ,
\label{bb.03}%
\end{equation}
where $\Phi\simeq\ln\phi$ is a new scalar field minimally coupled to gravity
and $\bar{\omega}=\bar{\omega}\left(  \omega_{BD}\right)  $. We observe that
the hyperbolic inflation is recovered when $F\left(  \Phi\right)  =const$. or
$F\left(  \Phi\right)  =e^{\bar{\kappa}\Phi}$, that is, $F\left(  \phi\right)
\simeq\phi^{\kappa}$.

In the hyperbolic inflation, for the scalar field potential $\bar{V}\left(
\Phi,\psi\right)  $ it holds, $\bar{V}_{,\psi}=0$ and $\bar{V}\left(
\Phi\right)  $ is an exponential function, thus in this work, for the
dynamical analysis we shall assume the power-potential $V\left(  \phi\right)
=V_{0}\phi^{\lambda}$. \ Let us focus in the case where $V\left(  \phi
,\psi\right)  =V\left(  \phi\right)  $.

We assume the isotropic and homogeneous background space described by the
spatially flat Friedmann--Lema\^{\i}tre--Robertson--Walker (FLRW) geometry
with line element%

\begin{equation}
ds^{2}=-dt^{2}+a^{2}\left(  t\right)  \left(  dx^{2}+dy^{2}+dz^{2}\right)  ,
\label{bd.05}%
\end{equation}
in which $a\left(  t\right)  $ is the scalar factor. The Hubble function is
defined as $H=\frac{\dot{a}}{a}$ in which $\dot{a}=\frac{da}{dt}$. The FLRW
geometry admits six isometries, the three translations and three rotations of
the three-dimensional flat hypersurface. Because we consider the scalar fields
to inherit the symmetries of the background space we find $\phi=\phi\left(
t\right)  $ and $\psi=\psi\left(  t\right)  $.

We derive the Ricciscalar
\begin{equation}
R=6\left[  \frac{\ddot{a}}{a}+\left(  \frac{\dot{a}}{a}\right)  ^{2}\right]  .
\label{bd.06}%
\end{equation}
By replacing in (\ref{bb.02}) the modified Friedmann's equations are derived%
\begin{equation}
-3H^{2}=3H\frac{\dot{\phi}}{\phi}-\frac{\omega_{BD}}{2}\left(  \frac{\dot
{\phi}}{\phi}\right)  ^{2}-\frac{1}{2}\frac{F^{2}\left(  \phi\right)  }{\phi
}\dot{\psi}^{2}-\frac{1}{\phi}V\left(  \phi\right)  ~, \label{bb.07}%
\end{equation}%
\begin{equation}
-\left(  3\phi H^{2}+2\phi\dot{H}\right)  =2H\dot{\phi}+\frac{\omega_{BD}%
}{2\phi}\dot{\phi}^{2}+\frac{1}{2}F^{2}\left(  \phi\right)  \dot{\psi}%
^{2}+\ddot{\phi}-V\left(  \phi\right)  \label{bb.08}%
\end{equation}

Furthermore, the equations of motion for the two scalar fields read
\begin{equation}
\omega_{BD}\left(  \ddot{\phi}-\frac{1}{2}\left(  \frac{\dot{\phi}}{\phi
}\right)  ^{2}+3H\dot{\phi}\right)  +6H^{2}\phi+\phi\left(  3\dot{H}+V_{,\phi
}-\frac{1}{2}\left(  F^{2}\right)  _{,\phi}\dot{\psi}^{2}\right)  =0~,
\label{bb.09}%
\end{equation}%
\begin{equation}
\ddot{\psi}+3H\dot{\psi}+\left(  \ln\left(  F^{2}\right)  \right)  _{,\phi
}\dot{\phi}\dot{\psi}=0. \label{bb.10}%
\end{equation}

The field equations (\ref{bb.07}) and (\ref{bb.08}) can be written in the
equivalent form%
\begin{equation}
3H^{2}=\rho_{eff~}~, \label{bb.11}%
\end{equation}%
\begin{equation}
2\dot{H}+3H^{2}=p_{eff}~, \label{bb.12}%
\end{equation}
where we have defined
\begin{equation}
\rho_{eff}=-3H\frac{\dot{\phi}}{\phi}+\frac{\omega_{BD}}{2}\left(  \frac
{\dot{\phi}}{\phi}\right)  ^{2}+\frac{1}{2}\frac{F^{2}\left(  \phi\right)
}{\phi}\dot{\psi}^{2}+\frac{1}{\phi}V\left(  \phi\right)  ~, \label{bb.13}%
\end{equation}%
\begin{equation}
p_{eff}=2H\frac{\dot{\phi}}{\phi}+\frac{\omega_{BD}}{2}\left(  \frac{\dot
{\phi}}{\phi}\right)  ^{2}+\frac{1}{2}\frac{F^{2}\left(  \phi\right)  }{\phi
}\dot{\psi}^{2}-\frac{\ddot{\phi}}{\phi}-\frac{1}{\phi}V\left(  \phi\right)
~. \label{bb.14}%
\end{equation}

\section{Exact solutions}

\label{sec4}

We investigate the existence of exact solutions of special interest for the
cosmological evolution. Indeed, we search for scaling solutions, $a\left(
t\right)  =a_{0}t^{p}$, exponential power-law $a\left(  t\right)
=a_{0}e^{H_{0}t}~$. For the analysis in this Section we consider an arbitrary
potential function $V\left(  \phi\right)  $.

Equation (\ref{bb.10}) gives the solution%

\begin{equation}
\dot{\psi}\left(  t\right)  =\psi_{0}a^{-3}\phi^{-2\kappa}\text{~}.
\label{bb.15}%
\end{equation}
where in the following we shall assume that $\psi_{0}$ is a non-zero constant,
otherwise we end to the usual Brans-Dicke theory.

For the power-law scale factor $a\left(  t\right)  =a_{0}t^{p}$, from
(\ref{bb.07}) and (\ref{bb.08}) with the use of (\ref{bb.15}) we end with the
second-order ordinary differential equation%
\begin{equation}
\ddot{\phi}+\omega_{BD}\frac{\dot{\phi}^{2}}{\phi}-\frac{p}{t}\dot{\phi}%
-\frac{2p}{t^{2}}\phi+t^{-6p}\psi_{0}^{2}\phi^{-2\kappa}=0. \label{bb.16}%
\end{equation}

A closed-form solution for the field equations is%
\begin{equation}
a\left(  t\right)  =a_{0}t^{p}~,~\phi=\phi_{0}t^{2\frac{1-3p}{1+2\kappa}%
}~,~\psi\left(  t\right)  =\frac{1+2\kappa}{1-2\kappa}\frac{\psi_{0}}%
{1-3p}\phi_{0}^{-2\kappa}t^{\frac{\left(  1-3p\right)  \left(  1-2\kappa
\right)  }{1+2\kappa}}~,
\end{equation}
with $\psi_{0}$ given by the expression
\begin{equation}
\psi_{0}=\pm\frac{\sqrt{2}\phi_{0}^{\frac{1}{2}+\kappa}}{\sqrt{1+4\kappa
+\kappa^{32}}}\sqrt{11p-21p^{2}-1+2\kappa-6p^{2}\kappa+4p\kappa^{2}%
-2\omega_{BD}\left(  1-6p+9p^{2}\right)  },
\end{equation}
and potential function%
\begin{equation}
V\left(  \phi\left(  t\right)  \right)  =\frac{\left(  1-2\kappa-p\left(
17+4\kappa\left(  3+\kappa\right)  +6p\left(  \kappa\left(  2\kappa-5\right)
-6\right)  \right)  \right)  }{\left(  1+2\kappa\right)  ^{2}}\phi_{0}%
t^{\frac{2\left(  3p+2\kappa\right)  }{1+2\kappa}}~,
\end{equation}
that is%
\begin{equation}
V\left(  \phi\right)  =\frac{\left(  1-2\kappa-p\left(  17+4\kappa\left(
3+\kappa\right)  +6p\left(  \kappa\left(  2\kappa-5\right)  -6\right)
\right)  \right)  }{\left(  1+2\kappa\right)  ^{2}}\phi_{0}^{\frac{1+2\kappa
}{1-3p}}\phi^{-\frac{3p+2\kappa}{1-3p}}.
\end{equation}

In the special case where $\kappa=-\frac{1}{2}$, it follows $p=\frac{1}{3}$,
$\psi_{0}=\sqrt{\frac{2\left(  1+2q_{1}\right)  -3q_{1}^{2}\left(
1-\omega_{BD}\right)  }{3}}$ and potential function~$V\left(  \phi\left(
t\right)  \right)  =-\frac{1}{2}\left(  q_{1}\left(  3q_{1}-10\right)
-4\right)  t^{-2+q_{1}}\phi_{0}$, that is,%
\begin{equation}
V\left(  \phi\right)  =-\frac{1}{2}\left(  q_{1}\left(  3q_{1}-10\right)
-4\right)  \left(  \frac{\phi}{\phi_{0}}\right)  ^{1-\frac{2}{q_{1}}}\phi_{0}.
\end{equation}

Moreover, the de Sitter universe is recovered when
\begin{equation}
\ddot{\phi}+\omega_{BD}\frac{\dot{\phi}^{2}}{\phi}-H_{0}\dot{\phi}%
+e^{-6H_{0}t}\psi_{0}^{2}\phi^{-2\kappa}=0. \label{bb.17}%
\end{equation}

We easily find that equation (\ref{bb.17}) admits an exponential closed-form
solution, thus, the field equations provide%
\begin{equation}
a\left(  t\right)  =a_{0}e^{H_{0}t}~,~\phi\left(  t\right)  =\phi_{0}%
e^{-\frac{6H_{0}}{1+2\kappa}t}~,~\psi\left(  t\right)  =\psi_{0}\phi
_{0}^{-2\kappa}e^{-3H_{0}\frac{1-2\kappa}{1+2\kappa}t},
\end{equation}
with
\begin{equation}
\psi_{0}=\pm\frac{\sqrt{6}H_{0}\sqrt{-\left(  7+2\kappa+6\omega_{BD}\right)
}}{1+2\kappa},
\end{equation}
and potential%
\begin{equation}
V\left(  \phi\left(  t\right)  \right)  =6H_{0}^{2}\frac{6+\kappa\left(
5-2\kappa\right)  }{\left(  1+2\kappa\right)  ^{2}}\phi_{0}e^{-\frac{6H_{0}%
}{1+2\kappa}t},
\end{equation}
that is%
\begin{equation}
V\left(  \phi\right)  =6H_{0}^{2}\frac{6+\kappa\left(  5-2\kappa\right)
}{\left(  1+2\kappa\right)  ^{2}}\phi.
\end{equation}

However, for $\kappa=-\frac{1}{2}$ we end with the closed-form solution
$\psi_{0}=0$ and $\phi\left(  t\right)  =\phi_{0}e^{\phi_{1}t}$ with linear
potential function. This solution is not of special interest because the
second scalar field, $\psi$ does not contribute in the cosmological fluid.

\section{Dynamical system analysis}

\label{sec5}

In order to perform the dynamical analysis we consider the new variables in
the $H$-normalization%
\begin{equation}
~x=\frac{\dot{\phi}}{\phi H}~,~y^{2}=\frac{V\left(  \phi\right)  }{3\phi
H^{2}}~,~z=\frac{\dot{\psi}F\left(  \phi\right)  }{\sqrt{6\phi}H}~,
\end{equation}
and%
\begin{equation}
\lambda=\phi\left(  \ln V\left(  \phi\right)  \right)  _{,\phi}~,~\kappa
=\phi\left(  \ln F\left(  \phi\right)  \right)  _{,\phi}~,~d\tau=Hdt~.
\end{equation}
Furthermore, we assume the exponential potential $V\left(  \phi\right)
=V_{0}e^{\lambda\phi}$ and the power-law function $F\left(  \phi\right)
=\phi^{\kappa}$, thus parameters $\lambda$ and $\kappa$ are constants. We
proceed with the isotropic background space.

For the spatially flat FLRW spacetime the field equations in the new variables
read%
\begin{equation}
1-\left(  \frac{3}{2\omega_{BD}}\right)  -\left(  \sqrt{\frac{\omega_{BD}}{6}%
}x-\frac{\sqrt{6}}{2\sqrt{\omega_{BD}}}\right)  ^{2}-y^{2}-z^{2}=0
\label{bb.20}%
\end{equation}%
\begin{align}
\left(  6+4\omega_{BD}\right)  \frac{dx}{d\tau}  &  =6\left(  1+\left(
3+2\lambda\right)  y^{2}+z^{2}\left(  4\kappa-3\right)  \right)  -x\left(
6\omega_{BD}+x\left(  7\omega_{BD}+6\right)  \right) \nonumber\\
&  +x\left(  \omega_{BD}\left(  1+\omega_{BD}\right)  x^{2}+6\left(
\lambda-\omega_{BD}\right)  y^{2}+6\left(  2\kappa+\omega_{BD}\right)
z^{2}\right)  ~, \label{bb.21}%
\end{align}%
\begin{align}
\frac{\left(  3+2\omega_{BD}\right)  }{y}\frac{dy}{d\tau}  &  =-\left(
3+4\omega_{BD}+\lambda\left(  3+2\omega_{BD}\right)  \right)  x+\omega
_{BD}\left(  1+\omega_{BD}\right)  x^{2}\nonumber\\
&  +6\left(  2+\omega_{BD}+\left(  \lambda-\omega_{BD}\right)  y^{2}+\left(
2\kappa+\omega_{BD}\right)  z^{2}\right)  ~, \label{bb.22}%
\end{align}%
\begin{align}
\frac{\left(  3+2\omega_{BD}\right)  }{z}\frac{dz}{d\tau}  &  =-\left(
3+4\omega_{BD}+\kappa\left(  6+4\omega_{BD}\right)  \right)  x+\omega
_{BD}\left(  1+\omega_{BD}\right)  x^{2}\nonumber\\
&  -6\left(  1+\omega_{BD}+\left(  \omega_{BD}-\lambda\right)  y^{2}\right)
+6\left(  2\kappa+\omega_{BD}\right)  z^{2}~. \label{bb.23}%
\end{align}

From equation (\ref{bb.20}) and for $\omega_{BD}>0$, variables $\left(
x,y,z\right)  $ take values on the sphere with radius $1+\left(  \frac
{3}{2\omega_{BD}}\right)  $ and center the point $\left(  \frac{\sqrt{6}%
}{2\sqrt{\omega_{BD}}},0,0\right)  $. However for $\omega_{BD}<0$, variables
$\left(  x,y,z\right)  $ are without boundaries.\ Thus for $\omega_{BD}<0$
Poincare variables should be used in order to perform a complete analysis.

At this point it is important to mention that the field equations are
invariant on the discrete transformations $y\rightarrow-y$ and $z\rightarrow
-z$ thus without loss of generality we select to work on the branch $y>0$.

From (\ref{bb.20}) we replace variable $y$ and we end with a two-dimensional
system on the variables $\left(  x,z\right)  $.

The stationary points $A=\left(  x\left(  A\right)  ,z\left(  A\right)
\right)  $ of the reduced system are
\[
A_{1}^{\pm}=\left(  -2,\pm\sqrt{-1-\frac{2}{3}\omega_{BD}}\right)
~,~A_{2}^{\pm}=\left(  \frac{3\pm\sqrt{9+6\omega_{BD}}}{\omega_{BD}},0\right)
~,
\]%
\[
A_{3}^{\pm}=\left(  \frac{6}{\lambda-2\kappa},\pm\frac{\sqrt{\lambda\left(
\lambda+5\right)  -3-2\kappa\left(  2+\lambda\right)  -6\omega_{BD}}}%
{\lambda-2\kappa}\right)  \text{~},
\]%
\[
A_{4}=\left(  \frac{2\left(  \lambda+2\right)  }{1-\lambda+2\omega_{BD}%
},0\right)  ~.
\]

Points $A_{1}^{\pm}$ describe asymptotic solutions where only the kinetic
parts of the two scalar fields contributes to the cosmological fluid, that is,
$y\left(  A_{1}^{\pm}\right)  =0$. The points are real and physically accepted
when~$\omega_{BD}>-\frac{3}{2}$, while the effective equation of state
parameter is $w_{eff}\left(  A_{1}^{\pm}\right)  =\frac{1}{3}\left(
1-4\kappa\right)  $. Thus, points $A_{1}^{\pm}$ describe acceleration when
$\kappa>\frac{1}{2}$.

Points $A_{2}^{\pm}$ are real when $\omega_{BD}>-\frac{3}{3}$ while they
provide $y\left(  A_{2}^{\pm}\right)  =0$, which means that only the kinetic
part of the scalar field $\phi$, contributes to the cosmological solution
Moreover, $w_{eff}\left(  A_{2}^{\pm}\right)  =\frac{6+3\omega_{BD}\pm
2\sqrt{9+6\omega_{BD}}}{3\omega_{BD}}$, that is, $w_{eff}\left(  A_{2}%
^{+}\right)  <-\frac{1}{3}$ for $-\frac{3}{2}<\omega_{BD}<0$.

For the stationary points $A_{3}^{\pm}$ we calculate $y\left(  A_{3}^{\pm
}\right)  =\frac{\sqrt{\left(  2\kappa-1\right)  \left(  2\kappa
-\lambda-3\right)  }}{\lambda-2\kappa}$and $w_{eff}\left(  A_{3}^{\pm}\right)
=1+\frac{2\left(  1+2\kappa\right)  }{\lambda-2\kappa}$. The points are real
when~$\left(  2\kappa-1\right)  \left(  2\kappa-\lambda-3\right)  >0$ and
$\omega_{BD}<\frac{1}{6}\left(  \lambda\left(  \lambda+5\right)
-3-2\kappa\left(  2+\lambda\right)  \right)  $. In these asymptotic solutions
all the components of the scalar fields contribute to the cosmological fluid.
Specifically, they can be seen as the set of the hyperbolic inflationary
solutions when $w_{eff}\left(  A_{3}^{\pm}\right)  <0$, in the Jordan frame.

Finally, point $A_{4}$ provides $y\left(  A_{4}\right)  =\frac{\sqrt
{5-\lambda\left(  4+\lambda\right)  -6\omega_{BD}}\sqrt{3+2\omega_{BD}}}%
{\sqrt{3}\left(  \lambda-1-2\omega_{BD}\right)  }$, $w_{eff}\left(
A_{4}\right)  =\frac{1+\lambda\left(  9+2\lambda\right)  -6\omega_{BD}%
}{3\left(  1-\lambda+2\omega_{BD}\right)  }$. The point is real for $\left(
5-\lambda\left(  4+\lambda\right)  -6\omega_{BD}\right)  \left(
3+2\omega_{BD}\right)  \geq0$. \ 

We proceed with the stability analysis of the stationary points so that we can
build the cosmological history.

The eigenvalues of the linearized system around points $A_{1}^{\pm}$ are
\begin{equation}
e_{1}\left(  A_{1}^{\pm}\right)  =1-2\kappa~,~e_{2}\left(  A_{1}^{\pm}\right)
=-2\left(  2\kappa-\lambda-3\right)  ~,
\end{equation}
which means that for $\left\{  \lambda<-2,\kappa>\frac{1}{2}\right\}  $ and
$\left\{  \lambda>-2,\kappa>\frac{3+\lambda}{2}\right\}  $ the stationary
points have negative eigenvalues which means that they are attractors. For the
stationary points $A_{2}^{\pm}$ we find
\begin{align}
e_{1}\left(  A_{2}^{\pm}\right)   &  =\pm\frac{\left(  1-2\kappa\right)
\left(  3+\sqrt{3\left(  3+2\omega_{BD}\right)  }\right)  }{2\omega_{BD}}~,\\
e_{2}\left(  A_{2}^{\pm}\right)   &  =\pm\frac{3\left(  1-\lambda+2\omega
_{BD}\right)  +\left(  1-\lambda\right)  \sqrt{3\left(  3+2\omega_{BD}\right)
}}{\omega_{BD}}~.
\end{align}
Hence, point $A_{2}^{+}$ is an attractor when $\left\{  \lambda\leq
-2,-\frac{3}{2}<\omega_{BD}<0,\kappa<\frac{1}{2}\right\}  $, $\left\{
-2<\lambda<1,\frac{\lambda^{2}+4\lambda-5}{6}<\omega_{BD}<0,\kappa<\frac{1}%
{2}\right\}  $ and $\left\{  \lambda>1,0<\omega_{BD}<\frac{\lambda
^{2}+4\lambda-5}{6},\kappa>\frac{1}{2}\right\}  $. On the other hand, point
$A_{2}^{-}$ is an attractor when $\left\{  \lambda\leq-5,-\frac{3}{2}%
<\omega_{BD}<0,\kappa<\frac{1}{2}\right\}  $, $\left\{  -5<\lambda
<-2,-\frac{3}{2}<\omega_{BD}<\frac{\lambda^{2}+4\lambda-5}{6},\kappa<\frac
{1}{2}\right\}  $.

For the stationary point $A_{4}$ the eigenvalues are
\begin{align}
e_{1}\left(  A_{4}\right)   &  =\frac{5-4\lambda-\lambda^{2}+6\omega_{BD}%
}{\lambda-1-2\omega_{BD}}~,\\
e_{2}\left(  A_{4}\right)   &  =\frac{3+4\kappa-5\lambda+2\kappa
\lambda-\lambda^{2}+6\omega_{BD}}{\lambda-1-2\omega_{BD}}.
\end{align}
It follows that the stationary point $A_{4}$ is an attractor when $\left\{
\lambda<-2,\omega_{BD}>\frac{\left(  \lambda-1\right)  \left(  5+\lambda
\right)  }{6},\kappa<\frac{5\lambda-3+\lambda^{2}-6\omega_{BD}}{2\left(
2+\lambda\right)  }\right\}  $, $\left\{  \lambda=-2,\omega_{BD}>-\frac{3}%
{2}\right\}  $, $\left\{  -2<\lambda\leq1,\omega_{BD}>\frac{\lambda-1}%
{2},\kappa>\frac{5\lambda-3+\lambda^{2}-6\omega_{BD}}{2\left(  2+\lambda
\right)  }\right\}  $ and $\left\{  \lambda>1,\omega_{BD}>\frac{\left(
\lambda-1\right)  \left(  5+\lambda\right)  }{6},\kappa>\frac{5\lambda
-3+\lambda^{2}-6\omega_{BD}}{2\left(  2+\lambda\right)  }\right\}  $.

\begin{figure}[ptb]
\centering\includegraphics[width=0.9\textwidth]{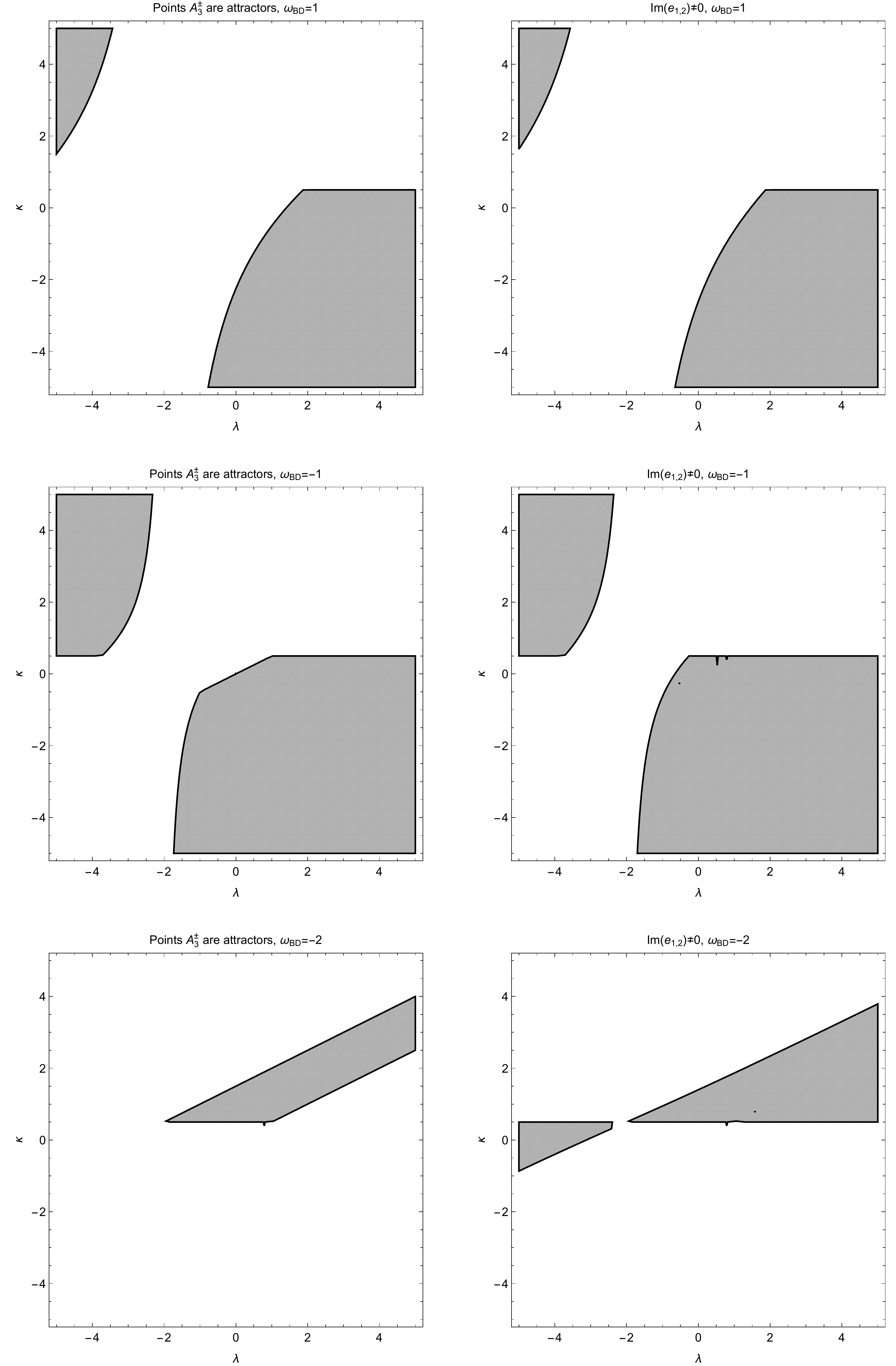}\caption{Region plots
of the real and imaginary parts of the eigenvalues $e_{1,2}\left(  A_{3}^{\pm
}\right)  $ in the plane $\left(  \lambda,\kappa\right)  $ for $\omega_{BD}%
=1$, $\omega_{BD}=-1$ and $\omega_{BD}=-2.$ Left column is the region where
the eigenvalues have negative real parts and right column is the region where
the imaginary parts of the eigenvalues are nonzero. }%
\label{fg1}%
\end{figure}

Finally, for the points $A_{3}^{\pm}$ \ the eigenvalues $e_{1,2}\left(
A_{3}^{\pm}\right)  $ are studied numerically. In Fig. \ref{fg1} we present
regions in the two-dimensional plane $\left(  \lambda,\kappa\right)  $ for
various values of the variable $\omega_{BD}$. What is important is that the
stationary points, when they are attractor, they are spirals, which is in
agreement with the hyperbolic inflation in the Einstein frame.

\begin{figure}[ptb]
\centering\includegraphics[width=1\textwidth]{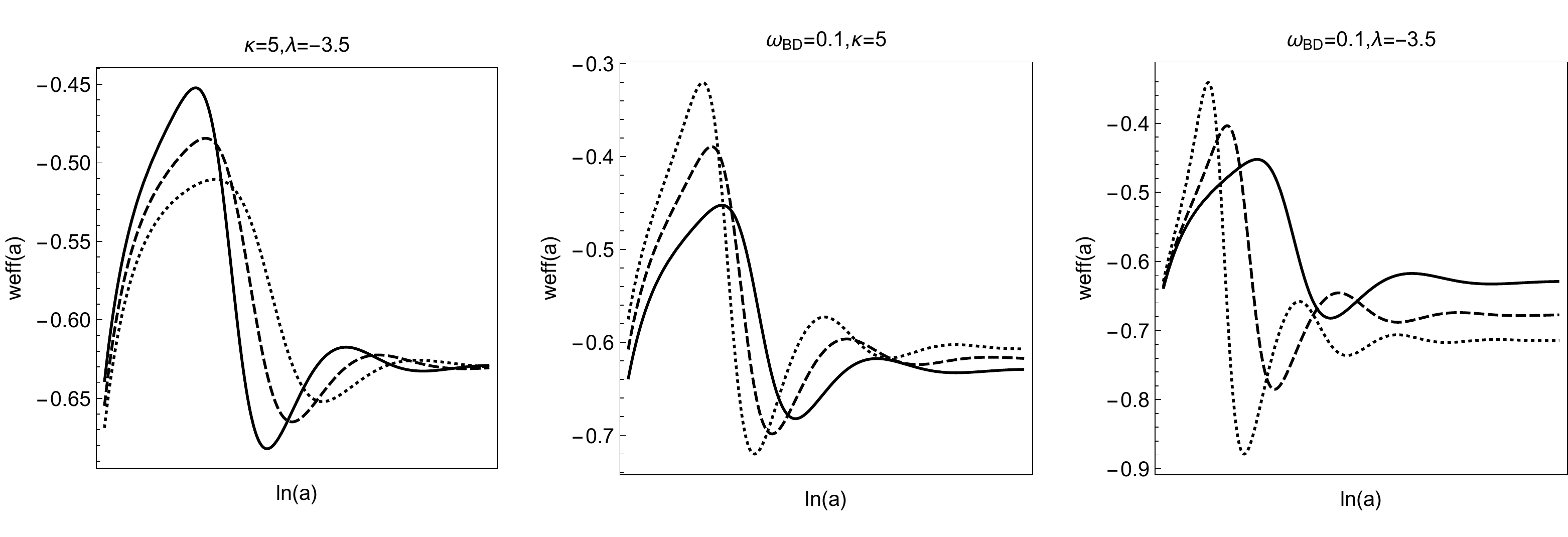}\caption{Qualitative
evolution of the $w_{eff}$ with a future atractor point $A_{3}^{+}$. Left Fig.
is for $\left(  \kappa=5,\lambda=-3.5\right)  $, and $\omega_{BD}=0.1$ (solid
line), $\omega_{BD}=0.2$ (dashed line) and $\omega_{BD}=0.3$ (dotted line).
Center Fig. is for $\left(  \omega_{BD},\kappa\right)  =\left(  0.1,5\right)
$, $\lambda=-3.5$ (solid line), $\lambda=-3.6$ (dashed line) and
$\lambda=-3.7$ (dotted line)\thinspace. Right line is for $\left(  \omega
_{BD},\lambda\right)  =\left(  0.1,-3.5\right)  $, $\kappa=-3.5$ (solid line),
$\kappa=-3.6$ (dashed line) and $\kappa=-3.7$ (dotted line)\thinspace.}%
\label{fg2}%
\end{figure}

\begin{figure}[ptb]
\centering\includegraphics[width=1\textwidth]{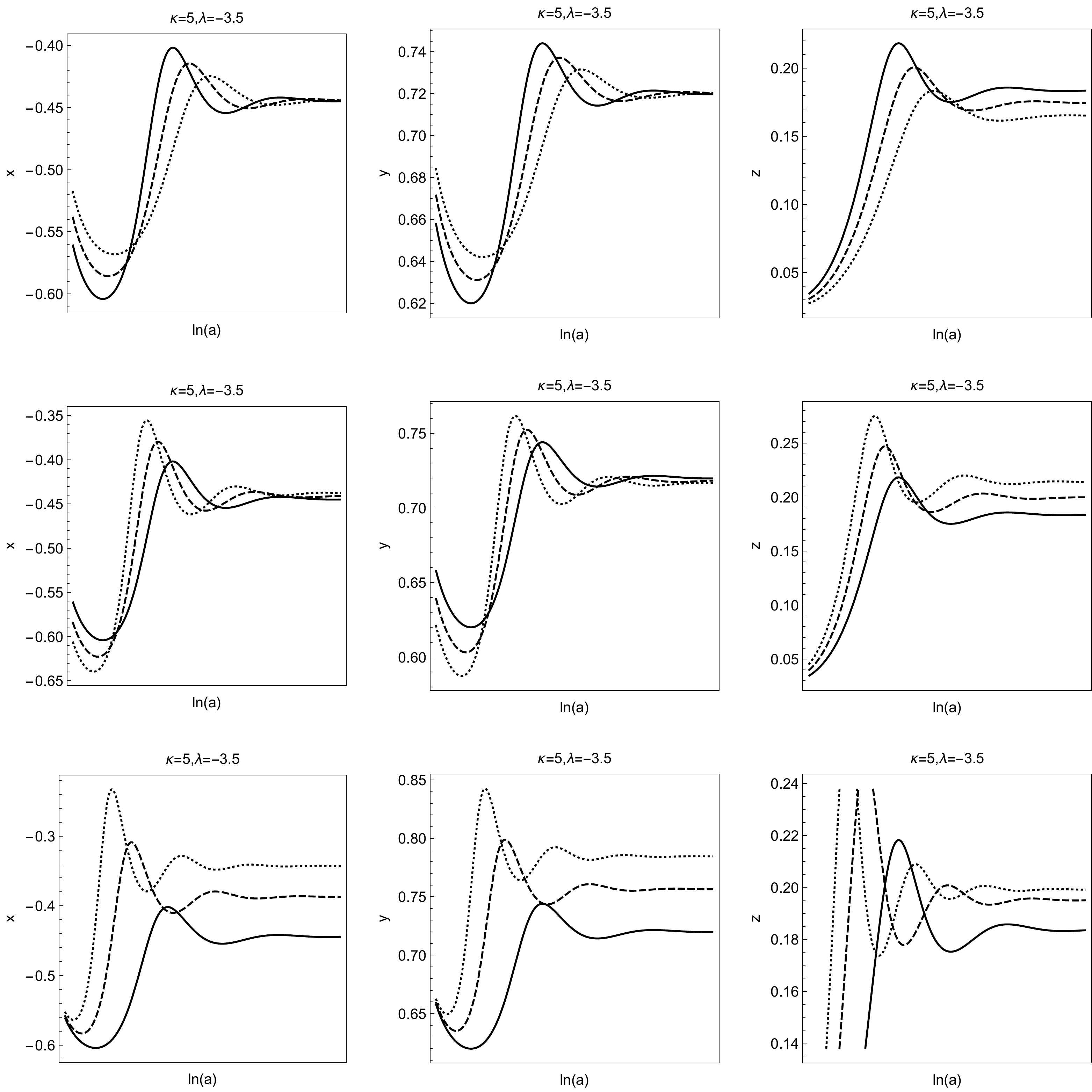}\caption{Qualitative
evolution of the dimensionless variables $\left(  x,y,z\right)  $ for various
values of the free parameters $\left(  \omega_{BD},\kappa,\lambda\right)  $.
First row is for $\left(  \kappa=5,\lambda=-3.5\right)  $, and $\omega
_{BD}=0.1$ (solid lines), $\omega_{BD}=0.2$ (dashed lines) and $\omega
_{BD}=0.3$ (dotted lines). Second row is for $\left(  \omega_{BD}%
,\kappa\right)  =\left(  0.1,5\right)  $, $\lambda=-3.5$ (solid lines),
$\lambda=-3.6$ (dashed lines) and $\lambda=-3.7$ (dotted lines)\thinspace.
Third row is for $\left(  \omega_{BD},\lambda\right)  =\left(
0.1,-3.5\right)  $, $\kappa=-3.5$ (solid lines), $\kappa=-3.6$ (dashed lines)
and $\kappa=-3.7$ (dotted lines)\thinspace.}%
\label{fg3}%
\end{figure}

In Fig. \ref{fg2} we present the qualitative evolution for the effective
equation of the state parameter for various values of the free parameters
$\left(  \omega_{BD},\lambda,\kappa\right)  $ where the hyperbolic inflation
is a future attractor. The qualitative evolution of the dimensionless
variables $\left(  x,y,z\right)  $ are presented in \ref{fg3}.

\subsection{Analysis at infinity}

For negative values of the Brans-Dicke parameter we define the new variables
\[
X=\frac{\rho}{\sqrt{1-\rho^{2}}}\cos\Theta~,~y=\frac{\rho}{\sqrt{1-\rho^{2}}%
}\sin\Theta\cos\Phi~,~z=\frac{\rho}{\sqrt{1-\rho^{2}}}\sin\Theta\sin\Phi
\]
where we have replaced $x=\sqrt{\frac{6}{\left\vert \omega_{BD}\right\vert }%
}X-\frac{3}{\left\vert \omega_{BD}\right\vert }$.

Assume now that we work on the region $\rho>1$, that is, \ in order $y\geq0$,
it follows $\Theta\in\lbrack0,\pi)$ and $\Phi\in(-\frac{\pi}{2},\frac{\pi}%
{2}]$

In the new variables the infinity is reached when $\rho\rightarrow1$. Thus,
with use of the use of the constraint equation, for $\rho\rightarrow1$ we end
with the following system of ordinary differential equations%
\begin{equation}
\dot{\rho}=0
\end{equation}%
\begin{equation}
\dot{\Phi}=-\frac{1}{2}\sqrt{\frac{3}{2\left\vert \omega_{BD}\right\vert }%
}\left(  2\kappa-\lambda\right)  \cos\Theta\sin\left(  2\Phi\right)
,~~\Theta=\frac{\pi}{4}\text{ or }\Theta=\frac{3\pi}{4}\text{.}%
\end{equation}
The first equation indicates that when the dynamical system lie at the
infinity it stays at $\rho\rightarrow1$. Moreover, from the second equation it
follows that $\Phi_{1}=0,~\Phi_{2}=\frac{\pi}{2}$.

Hence, the stationary points $B=\left(  \Theta\left(  B\right)  ,\Phi\left(
B\right)  \right)  $ are $B_{1}=\left(  \frac{\pi}{4},0\right)  $,
$B_{2}=\left(  \frac{\pi}{4},\frac{\pi}{2}\right)  $ , $B_{3}=\left(
\frac{3\pi}{4},0\right)  $ and $B_{4}=\left(  \frac{3\pi}{4},\frac{\pi}%
{2}\right)  $. For the points $B_{1}$ and $B_{3}$, we derive $z\left(
B_{1}\right)  =0$ and $z\left(  B_{3}\right)  =0$, while for the points
$B_{2}$ and $B_{4}$ we find $y\left(  B_{2}\right)  =0$ and $y\left(
B_{4}\right)  =0$.

As far as the stability properties of the points are concerned, we find that
points $B_{1}$ and $B_{4}$ are attractors for $2\kappa-\lambda>0$, otherwise
for $2\kappa-\lambda<0$ points $B_{2}$ and $B_{4}$ are attractors on the
surface $\rho\rightarrow1$.

\section{Conclusions}

\label{conc}

Hyperbolic inflation is a two-scalar field model in the Einstein frame where
the two scalar fields lie on a hyperbolic plane. This model admits an
attractor which describes an inflationary solution where the two scalar fields
contribute to the cosmological solution.

In this piece of work we consider a cosmological model of two-scalar fields in
the Jordan frame which recovers the multi-field model of hyperbolic inflation
under a conformal transformation. The main motivation of this study is to
investigate the nature of the attractor which corresponds to the hyperbolic
inflation under conformal transformations.

For our model, which is an extension of the Brans-Dicke and in the case of a
spatially flat FLRW geometry we performed a detailed analysis of the dynamics
of the field equations by using normal variables. The Brans-Dicke theory is
recovered, however we found a set of stationary points, namely $A_{3}^{\pm}$
where all the components of the gravitational Action Integral contribute to
the cosmological solution. The asymptotic solutions at the points $A_{3}^{\pm
}$ are scaling solutions which can describe accelerated universes. Thus these
solutions can be seen as the analogue of the hyperbolic inflation in Jordan frame.

Furthermore, surprisingly, points $A_{3}^{\pm}$ are stable spirals, which
means that they are in the same nature as the attractor of the hyperbolic
inflation in the Einstein frame. That is an interesting result because the
physical properties and the stability properties of the hyperbolic inflation
remain invariant under conformal transformation.

At this point, we wish to briefly discuss the hyperbolic inflationary solution
in the presence of anisotropy. We assume the locally rotational Bianchi I
spacetime with line element%
\begin{equation}
ds^{2}=-dt^{2}+a^{2}\left(  e^{2\sigma}dx^{2}+e^{-\sigma}\left(  dy^{2}%
+dz^{2}\right)  \right)  ~,
\end{equation}
and by using dimensionless-like variables as before, we conclude that the
isotropic inflationary solution described by the stationary points $A_{3}%
^{\pm}$ exist, with the same physical properties as before, and the same
stability properties. Hence, the hyperbolic inflation in the Jordan frame is
an isotropic attractor and can solve the isotropization problem as in the case
of the Einstein frame. Finally, we conclude that no new anisotropic exact
solutions are provided by this model except that of that of the standard
Brans-Dicke theory.

In a future work we plan to investigate in details the presence of curvature
and anisotropy in the initial conditions of the theory.

\end{document}